\documentclass{article}

\usepackage{arxiv}
\usepackage[utf8]{inputenc} 
\usepackage[T1]{fontenc}    
\usepackage{hyperref}       
\usepackage{url}            
\usepackage{booktabs}       
\usepackage{amsfonts}       
\usepackage{nicefrac}       
\usepackage{microtype}      
\usepackage{graphicx}
\usepackage{natbib}
\usepackage{doi}
\usepackage{amsmath}

\title{Approaching the conformal limit of quark matter with different chemical potentials}

\date{} 					

\author{ {Connor Brown} \\
	Center for Nuclear Research\\
	Kent State University\\
	Kent, OH 44242 USA \\
	\texttt{cbrow232@kent.edu} \\
	\And
	{Veronica Dexheimer} \\
	Center for Nuclear Research\\
	Kent State University\\
	Kent, OH 44242 USA \\
	\texttt{vdexheim@kent.edu} \\
        \AND
        {Rafael Bán Jacobsen} \\
	Center of Natural and Exact Sciences\\
	Universidade Federal de Santa Maria\\
	Santa Maria, RS, Brazil \\
	\texttt{rafaeljacobsen@gmail.com} \\
 	\And
	{Ricardo Luciano Sonego Farias} \\
	Center of Natural and Exact Sciences\\
	Universidade Federal de Santa Maria\\
	Santa Maria, RS, Brazil \\
	\texttt{ricardo.farias@ufsm.br} \\
}

\renewcommand{\undertitle}{ }
\renewcommand{\shorttitle}{Approaching the conformal limit of quark matter}

\begin{document}
\maketitle

\begin{abstract}
We study in detail the influence of different chemical potentials (baryon, charged, strange, and neutrino) on how and how fast a free gas of quarks in the zero-temperature limit reaches the conformal limit. We discuss the influence of non-zero masses, the inclusion of leptons, and different constraints, such as charge neutrality, zero-net strangeness, and fixed lepton fraction. We also investigate for the first time how the symmetry energy of the system under some of these conditions approaches the conformal limit. Finally, we briefly discuss what kind of corrections are expected from perturbative QCD as one goes away from the conformal limit.
\end{abstract}



\section{Introduction and Formalism}

In the zero temperature limit, baryons start to overlap at a few times saturation density and, through some mechanism that is not yet understood, quarks become effectively deconfined \cite{Baym:2017whm}. In this work we discuss dense matter in terms of baryon chemical potential $\mu_B$, instead of baryon (number) density $n_B$, as the former (together with other chemical potentials, such as electric charge $\mu_Q$ or strange $\mu_S$) is the fixed or independent quantity in the grand canonical ensemble. The correspondence between $n_B$ and $\mu_B$ is model dependent, but, at finite temperature, the $\mu_B$ at which deconfinement takes place is expected to be even lower (see e.g., \cite{Alford:2007xm}), which highlights the importance of studying quark matter. We are particularly interested in understanding the conformal limit, the asymptotically high $\mu_B$ at which matter can be described by a free (non-interacting) gas of massless quarks. For this reason, in the present work, we focus on modelling quark matter only and for the time being restrict ourselves to the zero-temperature limit. 

To describe the quarks, we make use of a free Fermi gas under different assumptions. To start, we describe them simply by a massless gas, then introduce different non-zero but constant quark masses, and vary independently the baryon, electric charge, and strange chemical potentials. We further link the chemical potentials by imposing charge neutrality and/or zero net strangeness. We also discuss the role played by leptons, discussing beta equilibrium and the role played by neutrinos (with chemical potential $\mu_\nu$). We investigate large $\mu_B$ and different $\mu_Q$ and $\mu_\nu$, as these are important for astrophysical scenarios, such as neutron stars and neutron-star mergers. On the other hand, we investigate the effects of $\mu_S$, which is important for discussions related to relativistic heavy-ion collisions and the early universe \cite{Letessier:1993hi}.

We also discuss the symmetry energy of quark matter for some of the constraints we study and investigate how it changes as we approach the conformal limit. The so called symmetry energy (which is really the \emph{asymmetry energy}) is one of the most important features of nuclear physics in general, since it is related to the ratio between the different components of the nuclear systems \cite{Baldo:2016jhp}. Several works have addressed the symmetry energy of quark matter \cite{Chu:2012rd,Chen:2017och,Wu:2018kww,Thakur:2017dtz}. This physical quantity is defined as the difference of energy per baryon $E/N_B$ (or energy density per baryon density ${\varepsilon}/{n_B}$) between completely isospin asymmetric matter $\delta=1$ and isospin-symmetric matter $\delta=0$: 
\begin{equation}
   E_{\rm{sym}} = \frac{ E_{\delta=1}}{N_B} - \frac{E_{\delta=0}}{N_B} = \frac{\varepsilon_{\delta=1}}{n_B} - \frac{\varepsilon_{\delta=0}}{n_B}\ ,
\end{equation}
where $\delta$ was originally defined for matter with neutrons and protons in terms of densities $n_i$ as 
\begin{equation}
\delta = \frac {n_n-n_p}{n_n+n_p}\ .
\label{2}
\end{equation}
In this case and also when one is considering up and down quarks, $\delta$ can also be written as
\begin{equation}
\delta = -2Y_{I}=1-2 Y_Q\ ,
\label{3}
\end{equation}
for non-strange matter, using the Gell-Mann–Nishijima formula \cite{Nakano:1953zz} with fractions $Y_I$ and $Y_Q$ summing over $i=$ baryons or quarks and defined in terms of particle isospin $I_i$ and electric charge $Q_i$, respectively
\begin{equation}
\rm{ Y_I = \frac{\sum_i I_i n_i}{\sum_i n_i}}\ ,\quad Y_Q = \frac{\sum_i Q_i n_i}{\sum_i n_i}\ ,
\label{4}
\end{equation}
with baryon (number) density $n_B=\sum n_i$, where the quark densities $n_i$ are divided by 3.

However, it is important to note that, as discussed in Ref.~\cite{Aryal:2020ocm} and Appendix A of Ref.~\cite{Yao:2023yda}, in the presence of hyperons (or in our case strange quarks), Eq.~\eqref{3} does not apply. For this reason, we restrain to the discussion of symmetry energy for the 2-flavor case (with up and down quarks).

When leptons are included, we assume beta equilibrium, in which case electrons and muons have chemical potential $\mu_e=\mu_\mu=-\mu_Q$. In the special case that (electron and muon) neutrinos are trapped, $\mu_\nu$ is determined by fixing the lepton fraction
\begin{equation}
Y_l = \frac{\sum_{lep} n_{lep}}{\sum_i n_i}\ ,
\label{5}
\end{equation}
usually hold equal to the canonical value $0.4$, to simulate conditions created in supernova explosions \cite{Burrows:1986me}.

Finally, we briefly discuss the effects of interactions in the case that they are week enough to be discussed perturbatively, i.e., using perturbative Quantum Chromodynamics, pQCD). At large temperatures and/or quark chemical potentials, the strong coupling becomes small enough to allow an infinite number of terms to be approximated by a finite number of terms to describe interactions \cite{Politzer:1973fx}. At zero temperature, QCD needs perturbative expansions and normalization group techniques. The problem was first addressed in the late seventies \cite{Freedman:1976xs,Freedman:1976dm,Freedman:1976ub} showing that already at the second order in  $\alpha_S$ log-singularities and non-trivial effects due to the renormalization scale appear. In the particular case of beta equilibrium, first-order corrections cancel out, leading to the very simple and popular description for
neutron stars in terms of free quarks plus a constant bag correction \cite{Alcock:1986hz}. This description is consistent with our work, up to a bag constant (usually fitted to phenomenology). But, even in the massless beta equilibrium
case, it has been shown that the effect of interactions is not negligible in the density or chemical potential regime relevant for astrophysics \cite{Fraga:2001id}.

At zero temperature, pQCD corrections have been calculated up to next-to-next-to-next-to-leading order (N$^3$LO) \cite{Gorda:2021kme,Gorda:2021znl} (see Eq.~42 of \cite{Vuorinen:2024qws}, where our free calculations would correspond to the term of order zero in the coupling $\alpha_s$). With non-zero quark masses, pQCD corrections have been calculated up to next-to-next-to-leading order (N$^2$LO) \cite{Fraga:2004gz,Kurkela:2009gj,Graf:2015tda,Gorda:2021gha}. In this case, the QCD running coupling constant implies running non-zero masses at finite chemical potentials.
See recently published lecture notes of pQCD in the context of astrophysics for more details \cite{Vuorinen:2024qws}. In this work, we do not study interactions, but simply discuss different kinds of conformal limit (for free quarks) and quantify how different they are from each other. Nevertheless, even in this simplest scenario, as a first attempt to clarify, e.g., the influence of various chemical potentials on the conformal limit, our approach is complementary to the study of interactions.

\section{Results}

We describe in detail the free Fermi gas formalism we use in this work (for quarks and leptons) in Appendix A. We begin our discussion by ignoring the contribution of leptons to the thermodynamical quantities (later we include different possibilities and discuss them). In the figures that follow, the pressure $P$ and baryon density $n_B$ are normalized by respective values of a free gas with the same number of quark flavors included, but with quark masses $m_i=0$ and $\mu_Q=\mu_S=0$. Simple analytical equations for the pressure of all the massless cases discussed in this work are derived in Appendix B. We start our discussion considering only one chemical potential, and then expand our discussion to two and three chemical potentials.

\subsection{One chemical potential $\mu_B$}

\begin{figure}
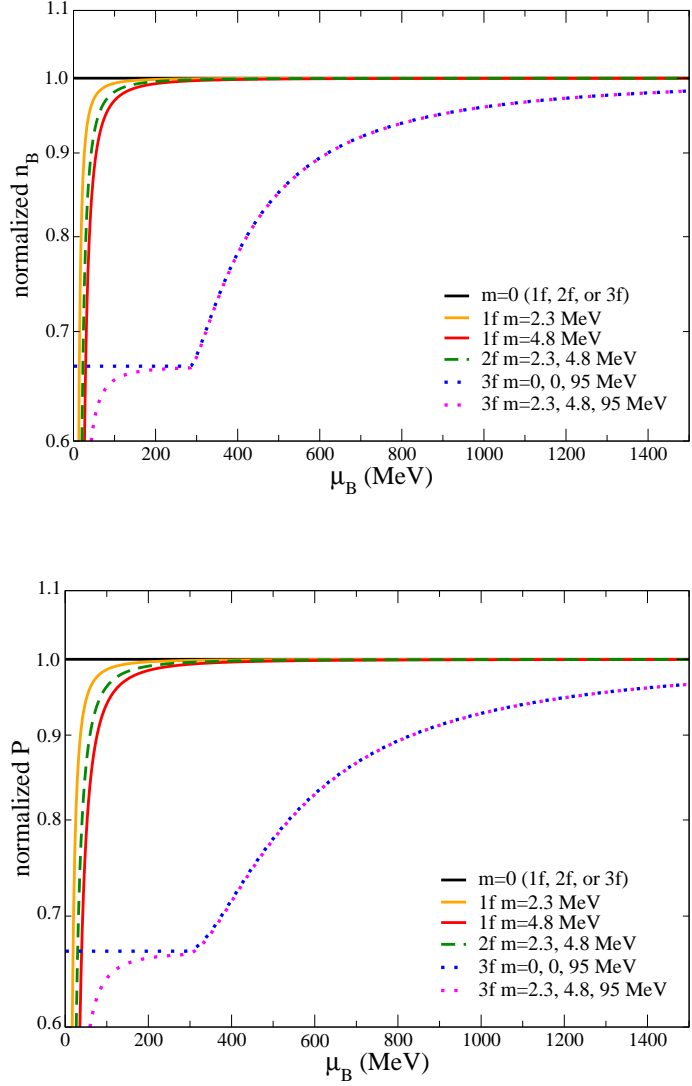

\setlength{\lineskip}{32pt}
\centering
    \includegraphics[width=.55\textwidth]{fig1a.eps}  
    \includegraphics[width=.55\textwidth]{fig1b.eps}
    \caption{Baryon density (upper panel) and pressure (lower panel) of quarks with different number of flavors and different masses normalized by the respective massless cases.}
    \label{fig1}
\end{figure}

We start by comparing the quark mass effect on $n_B$ versus $\mu_B$  in the left upper panel of Fig.~\ref{fig1}. Because in this case $\mu_Q$ and $\mu_S$ are zero, all quarks present the same chemical potential $\mu_i=\mu_{\rm{u}}=\mu_{\rm{d}}=\mu_{\rm{s}}=\frac 1 3\mu_B$. Due to our normalization (thermodynamical quantities divided by the massless case with the respective number of flavors), all massless cases have constant value $1$. Nevertheless, this does not mean that they are the same (if not normalized). To discuss the effect of quark masses, we start with 1 flavor with mass of the up $m=2.3$ MeV or down $m=4.8$ MeV quarks, then we look at the 2-flavor case with these masses for both light quarks. After that, we look at 3-flavors and use first only non-zero mass for the strange quark $m=95$ MeV and then the masses for the 3 quarks. The quark masses we use correspond to the Particle Data Group (PDG \cite{ParticleDataGroup:2014cgo}, within the error bars of updated values \cite{Workman:2022ynf}). From hereon we refer to these masses as “realistic”.

We find that the introduction of realistic quark masses decreases the density for low $\mu_B$, with the s-quark mass affecting the density until larger $\mu_B$ (up to $621$ MeV) than the two light quarks (up to $55$ MeV). To calculate these thresholds, we use throughout this paper the criteria of a deviation of $10\%$ from the black line with value 1. 
For $P$ versus $\mu_B$, shown in the lower panel of Fig.~\ref{fig1}, the lines are very similar in shape (to the ones in the upper panel of the figure). The introduction of realistic quark masses decreases again $P$ for low $\mu_B$, with the s-quark mass affecting the pressure until larger $\mu_B$ (up to $834$ MeV) than the two light quarks (up to $77$ MeV).

\begin{figure}
\setlength{\lineskip}{29pt}
\centering
    \includegraphics[width=.52\textwidth]{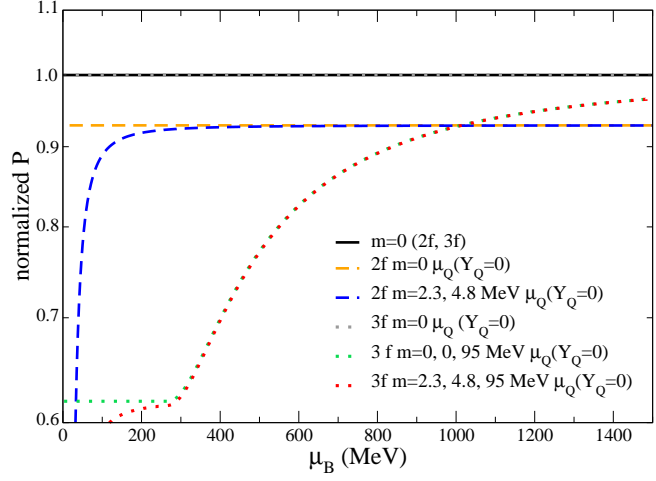}
    \includegraphics[width=.54\textwidth]{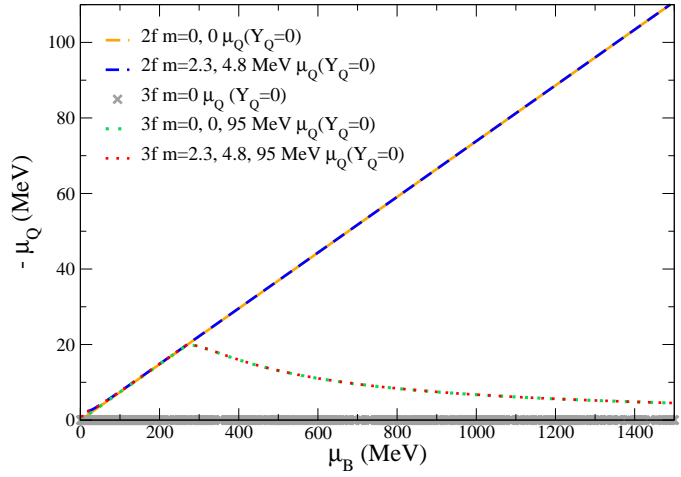} 
    \includegraphics[width=.52\textwidth]{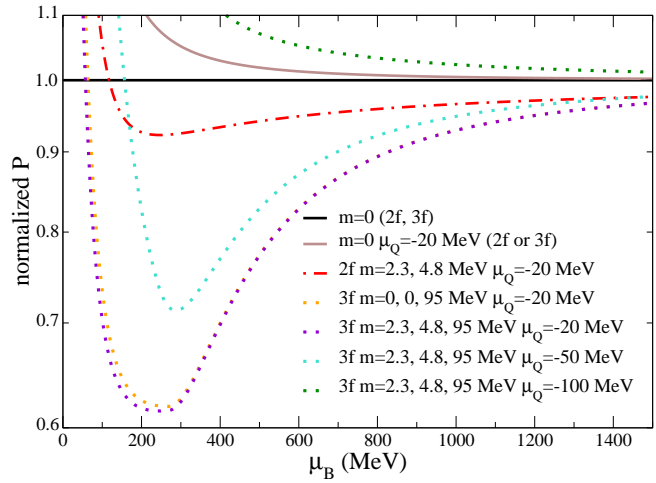}
    \caption{Pressure (upper panel) and electric charge chemical potential (middle panel) of quarks with 2 chemical potentials normalized by the respective massless case with one chemical potential, $\mu_B$. The electric charge  chemical potential is determined by charge neutrality. For massless 3-flavor quarks, the cases with and without $\mu_Q$ coincide. Lower panel: Pressure of quarks with 2 chemical potentials, being $\mu_Q$ fixed to different values, normalized by the respective massless case with one chemical potential, $\mu_B$.}
    \label{fig2}
\end{figure}

\subsection{Two chemical potentials $\mu_B$ and $\mu_Q$}

Now, we abandon the unphysical 1-flavor case, and continue with 2- and 3-flavor cases. The 2-flavor case has recently become more relevant for dense matter because it has been shown that the core of neutron stars can harbor 3-, as well as 2-flavor quark matter \cite{Holdom:2017gdc}. For this case we add another (electric charge) chemical potential, breaking some of the degeneracy in the quark chemical potentials: $\mu_{\rm{up}}=\frac 1 3 \mu_B + \frac 2 3 \mu_Q$, 
$\mu_{\rm{down}}=\mu_{\rm{strange}}=\frac 1 3 \mu_B - \frac 1 3 \mu_Q$. Once more, we normalize thermodynamical quantities dividing by the respective values of the same quantity for a free gas with the same number of quark flavors included, but with $m_i=0$, in addition to $\mu_Q=0$. Following this procedure, we aim at determining how the conformal limit and its deviation depend on $\mu_Q$.

When $\mu_Q$ is determined by charge neutrality, the results even for the massless case depend on the number of flavors. In this case, only the 3-flavor case is coincidentally equal to the $\mu_Q=0$ case (see the explanation following Eqs. \eqref{a37} to \eqref{a40} in Appendix B). For 2-flavor, this is not the case, and the pressure is lower than in the $\mu_Q=0$ case, establishing a new lower conformal limit (see upper panel of Fig.~\ref{fig2}). Expressions for the pressure for each particular chemical potential case (always keeping $m_i=0$ for simplicity) can be found in Appendix B. Compare e.g., Eqs. \eqref{17} and \eqref{32}.
When adding quark masses, $\mu_Q$ determined by charge neutrality lowers the pressure (in comparison to the respective massless case and to the massless case with $\mu_Q=0$) such that it goes to the respective conformal limit at larger $\mu_B$. Using again the criteria of $10\%$ deviations from the respective conformal limit, the s-quark mass affects pressure until $\mu_B=839$ MeV and the two light quark masses until $\mu_B=118$ MeV.

Nevertheless, one issue about this approach should be noted: we are comparing very small values of $\mu_Q$  with very large values of $\mu_B$. See the middle panel of Fig.~\ref{fig2} for a comparison. This is particularly the case for 3-flavors of quarks, and (except for extremely low $\mu_B$) this behavior is independent of the quark masses. For small values of $\mu_B$, both for 2 and 3-flavors, the dependence of $\mu_Q$ and $\mu_B$ can be predicted in fair agreement with Eq.~\eqref{eqmus2}.
%
For this reason, next, we add a fixed electric charge chemical potential to study how it affects the conformal limit, which translates into an increase in pressure (see, e.g., the different lines for 3-flavor quark matter with realistic masses in the lower panel of Fig.~\ref{fig2}),  specially at low values of $\mu_B$. For massless quarks and $\mu_Q=-20$ MeV, the pressure is always above the conformal limit for $\mu_Q=0$, independently of the number of flavors. Once the quark masses are finite, the pressure decreases, specially in the 3-flavor case (but also for the 2-flavor case). For larger absolute values of $\mu_Q$, the pressure becomes larger, even going above the conformal case (with and without $\mu_Q$). For example, for the 3-flavor case with realistic quark masses and $\mu_Q=-50$ MeV, the pressure deviates $10\%$ (of the $\mu_Q=0$ conformal limit) at $\mu_B=698$ MeV and for $\mu_Q=-100$ MeV at $\mu_B=415$ MeV (the latter one from above). Finally, there is one important remark regarding the behavior of the normalized pressure: in the lower panel of Fig.~\ref{fig2}, it is shown that this physical quantity decreases for small values of $\mu_B$; however, this behavior doesn't mean that the pressure itself (not normalized) is not a monotonically increasing function of $\mu_B$. Here, we must remember that our normalization is carried out by dividing the thermodynamical quantities (such as pressure) by the massless case with the respective number of flavors, and the free Fermi pressure of this system of massless quarks used for normalization scales as $\mu_B^4$; therefore, in those ranges of $\mu_B$ where $P$ for massive quarks increases at a lower rate than $\mu_B^4$, the normalized pressure decreases without implying any thermodynamical inconsistency.

\subsection{Three chemical potentials $\mu_B$, $\mu_Q$, and $\mu_S$ or $\mu_\nu$}

\begin{figure}
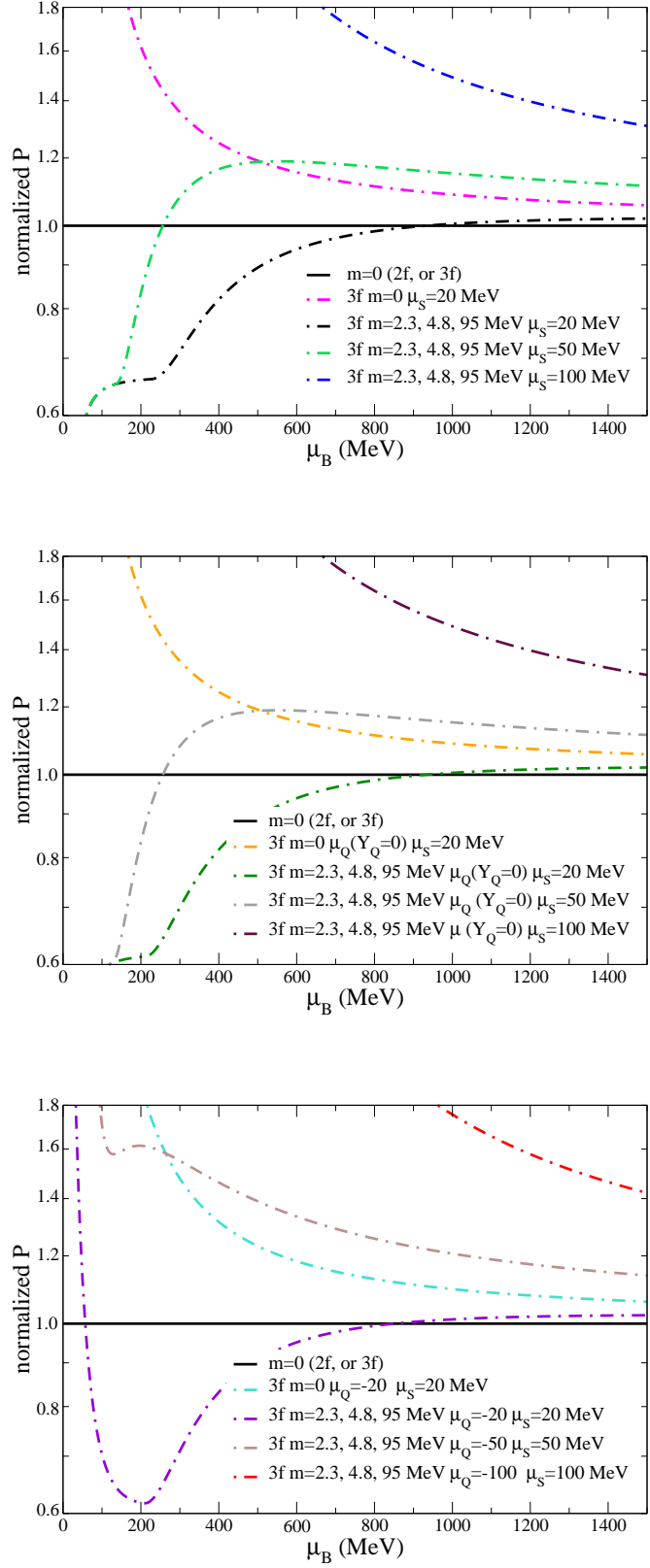

\setlength{\lineskip}{32pt}
\centering
    \includegraphics[width=.53\textwidth]{fig3a.eps}
    \includegraphics[width=.53\textwidth]{fig3b.eps}
    \includegraphics[width=.53\textwidth]{fig3c.eps}
    \caption{Pressure of quarks with 2 or 3 chemical potentials, including the strange chemical potential, normalized by the respective massless case (with one chemical potential, $\mu_B$). The electric charge chemical potential is either zero (upper panel), determined by charge neutrality (middle panel), or fixed (lower panel). For massless 3-flavor quarks, the cases with charge neutrality and without $\mu_Q$ coincide.}
    \label{fig3}
\end{figure}

Going further, we can add another (strange) chemical potential and constrain it, e.g., to strangeness neutrality. The issue is that at zero temperature strangeness neutrality means that there are no strange quarks, and the 3-flavor reduces to the 2-flavor case. For this reason, we fix $\mu_S$ instead to specific values. $\mu_S$ breaks the degeneracy in the remaining quark chemical potentials: $\mu_{\rm{up}}=\frac 1 3 \mu_B + \frac 2 3 \mu_Q$, 
$\mu_{\rm{down}}=\frac 1 3 \mu_B - \frac 1 3 \mu_Q$, $\mu_{\rm{strange}}=\frac 1 3 \mu_B - \frac 1 3 \mu_Q + \mu_S$. Once more, we normalize thermodynamical quantities dividing by the respective values of the same quantity for a free gas with the same number of quark flavors included, but with $m_i=0$, in addition to $\mu_Q=0$. 

Fixing $\mu_S$ increases the pressure, similar to fixing $\mu_Q$. Compare, for example, the massless 3-flavor case in the upper panel in Fig.~\ref{fig3} and lower panel in Fig.~\ref{fig2} and note that the pressure for a given $\mu_B$ is now much higher. When quark masses are added, the similarity disappears, because $\mu_S$ only affects the strange quarks, which do not appear for low values of $\mu_B$, unless the $\mu_S$ value is larger than the strange quark mass, which corresponds to our case of $\mu_S=100$ MeV. For $\mu_S= 50$ and $\mu_S= 100$ MeV, the $10\%$ deviation from the conformal limit takes place at $\mu_B=1743$ and $\mu_B=4227$ MeV, respectively (both from above). 

Now we consider the case in which additionally $\mu_Q\neq 0$, determined to reproduce charge neutrality (middle panel of Fig.~\ref{fig3}). For massless 3-flavor quarks, the cases with charge neutrality and without $\mu_Q$ coincide. When masses are introduced, the curves are still very similar (to the upper panel for the $\mu_Q=0$ case), except at very small $\mu_B$, where the quark masses are comparable to both $\mu_B$ and $\mu_Q$. For $\mu_S= 50$ and $\mu_S= 100$ MeV, the $10\%$ deviation from the conformal limit takes place at $\mu_B=1743$ and $\mu_B=4227$ MeV, respectively (both from above). 
When a fixed value of $\mu_Q$ is used, it increases the pressure further, specifically at low $\mu_B$ (see lower panel of Fig.~\ref{fig3}). For $\mu_Q=\mu_S= 50$ and $\mu_Q=\mu_S= 100$ MeV, the $10\%$ deviation from the conformal limit takes place at $\mu_B=2070$ and $\mu_B=4723$ MeV, respectively (both from above).

\begin{figure}
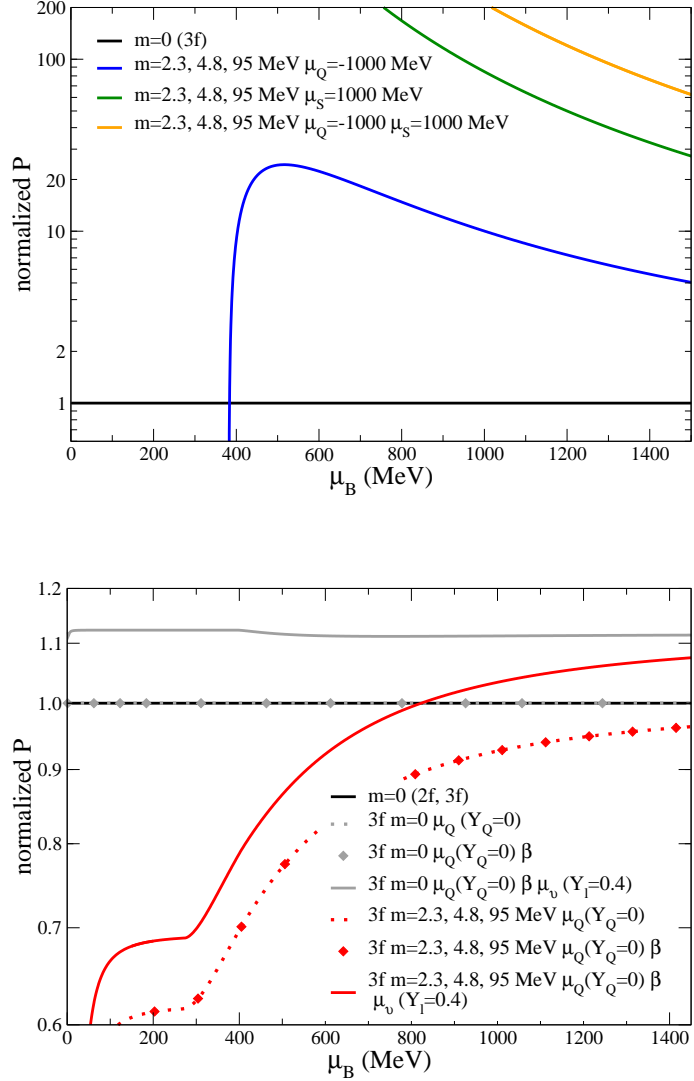

\setlength{\lineskip}{32pt}
\centering
    \includegraphics[width=.55\textwidth]{fig4a.eps}
    \includegraphics[width=.55\textwidth]{fig4b.eps}
    \caption{Upper panel: Pressure of quarks with 2 or 3 large chemical potentials, normalized by the respective massless case with one chemical potential, $\mu_B$. Lower panel: Pressure of quarks and leptons with 2 or 3 chemical potentials, normalized by the respective massless case with one chemical potential, $\mu_B$. For beta equilibrium with leptons, $\mu_Q$ is determined by charge neutrality. When neutrinos are present, their chemical potential $\mu_\nu$ is determined by fixing the lepton fraction, $Y_l$.}
    \label{fig4}
\end{figure}

Next, we investigate the effects of having much larger values of $\mu_Q$ and $\mu_S$, comparable to $\mu_B$, for 3 flavors of quarks in the upper panel of Fig.~\ref{fig4}. As expected, the changes due to the additional chemical potentials take place at much lower $\mu_B$ (notice the different scale in the y-axis of the figure) and practically all the curves are above the one chemical potential ($\mu_B$) conformal limit. An exception is the case with large (negative) $\mu_Q$ (and $\mu_S=0$) because, according to Eqs.~\ref{mu} and \ref{e}, quarks can only exist after a given $\mu_B=381$ MeV, at which the momentum $k_i$ and $P$ become finite (see Eq.~\ref{previous} for the massless case). In this case, the pressure differs from the one chemical potential conformal limit by more than $10\%$ until $\mu_B=10~583$ MeV. In the case of large $\mu_S$, quarks can exist at any $\mu_B$ and the pressure differs from the one chemical potential conformal limit by more than $10\%$ until $\mu_B=44~237$ MeV. When we combine large $\mu_S$ and (absolute value of) $\mu_Q$, the pressure differs from the one chemical potential conformal limit by more than $10\%$ until $\mu_B=48~897$ MeV. In this case, the curve in the upper panel of Fig.~\ref{fig4} begins only at $\mu_B=1000$ MeV. This can be understood once more from Eqs.~\ref{mu} and \ref{e}. The same effect can also be seen (although more subtle) in the bottom panel of Fig.~\ref{fig2}, where the fixed $\mu_Q$ cases start at $\mu_B=-\mu_Q$.

Finally, we investigate changes due to the inclusion of a free gas of leptons (electrons and muons) in beta equilibrium (and participating in the fulfillment of charge neutrality). As it can be seen in the lower panel of Fig.~\ref{fig4}, the inclusion of leptons (which appear in very small numbers or not at all) doesn't change the pressure. The picture changes though when lepton fraction is fixed. In this case, which also includes neutrinos, the pressure is considerably higher, not because of the neutrinos themselves, but because the larger amount of negative leptons forces the appearance of a large amount of up quarks, changing considerably the quark composition of the system and the stiffness of the equation of state \cite{Jimenez:2020vyu}. This same stiffening occurs with nucleons in equilibrium with a fixed fraction of leptons in the context of protoneutron stars and supernova explosions (see, e.g., \cite{PhysRevC.98.055805} and \cite{1980ApJ...238..717G}). The grey full line shows a kink for $\mu_B\sim400$ MeV, when the muons appear. Note that the difference in massless versus massive quarks is still very pronounced when $Y_l$ is fixed.

\subsection{Symmetry energy}

As already discussed, we calculate the symmetry energy only for the 2-flavor case, for which it was originally defined. The symmetry energy can be defined for strange matter, but the problem in this case is that it becomes ambiguous, as the two sides that appear in Eq.~\ref{3} (in terms of $Y_I$ or $Y_Q$) become different because of the  Gell-Mann–Nishijima formula. Instead of choosing one particular definition for strange matter, we prefer not to use it. For more details regarding the treatment of the symmetry energy for strange matter, see, e.g., \cite{Chu:2012rd}.

\begin{figure}
\centering
    \includegraphics[width=.63\textwidth]{fig5.eps}
    \includegraphics[width=.82\textwidth]{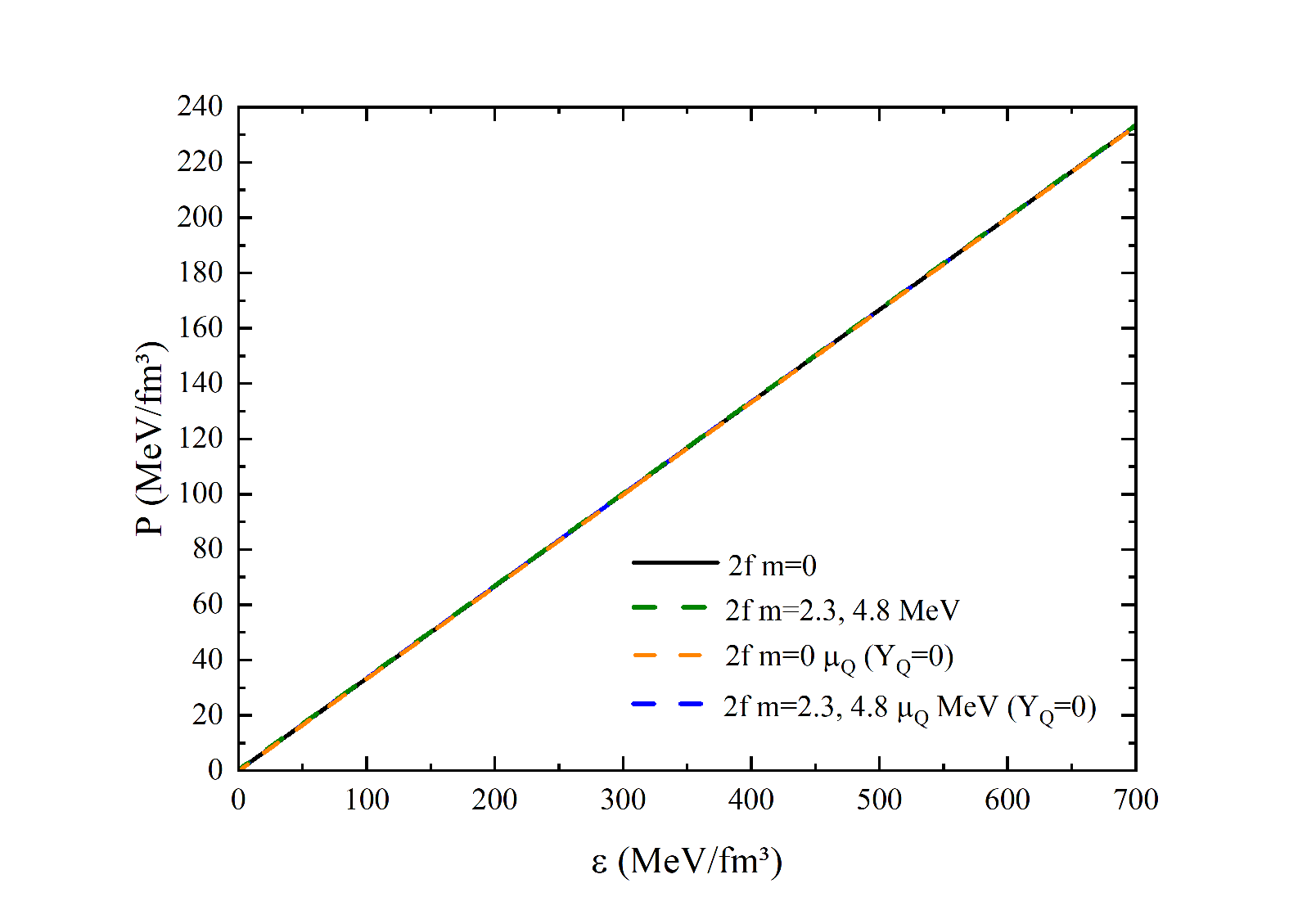}
    \caption{Upper panel: Symmetry energy of 2 flavor of quarks as a function of baryon density for different masses. The two curves overlap. 
    Lower panel: Equation of state (pressure vs. energy density) for the two extreme cases subtracted in the panel above, isospin-symmetric matter $\delta=0$ and charge neutral matter $\delta=1$, shown for realistic or zero quark masses. In this scale, the four curves overlap.}
    \label{fig5}
\end{figure}

To perform the calculation, we fix $n_B$ in this case (instead of $\mu_B$ as we have been doing) because the symmetry energy is defined for a given $n_B$, but limit the x-axis to approximately the corresponding range from the previous figures. The upper panel of Fig.~\ref{fig5} shows that the curves are a monotonically increasing function of density and that the light quark masses don't affect the results. Indeed, the effect of the mass is expected to be negligible in the symmetry energy of a free quark gas, since the quark masses are taken as very small at any density, while physically they should increase as density decreases. At very high density, much above the range shown in Fig.~\ref{fig5}, when the interactions are so weak that they can be neglected, the independence observed on the quark mass means that it is correct to consider the conformal (or massless) limit for the high density limit of QCD, because massless quarks or quarks with physical masses of the order of the $MeV$ are basically equivalent. This feature is reinforced by the clear overlapping of the equations of state $P(\varepsilon)$ for the corresponding 2-flavor cases here analyzed, as shown in the lower panel of Fig.~\ref{fig5}. In this plot, the range of $\mu_B$ is the same of the previous figures (running form $0$ to, approximately, $1400$ MeV). Notice that the independence on the light quark masses applies to every thermodynamical quantity that is not normalized by the respective conformal limit (and does not include derivatives). Numerically, we define $\delta=0$ as the 2-flavor $\mu_Q=0$ case (corresponding to the 2-flavor lines in Fig.~\ref{fig1}) and 
$\delta=1$ as the 2-flavor $Y_Q=0$ case (with $\mu_Q\neq0$ corresponding to the 2-flavor lines in the top and middle panels of Fig.~\ref{fig2}).

\section{Discussion and Conclusions}

Perturbative corrections to a free gas of quarks due to interactions always bring down the pressure to lower values. Although these corrections have been calculated to higher orders for massless and massive (strange) quarks, they cannot directly be carried out to low baryon chemical potentials $\mu_B$ (or, interchangeably, low baryon densities $n_B$ in the zero-temperature limit). This steams from the fact that matter in this regime is not perturbative and at some point a phase transition to hadronic matter takes place. As a result, for the relevant regime of densities inside neutron stars, $\mu_B\leq1500$ MeV, pQCD predicts that the pressure is lower than $80\%$ of the free gas value (see for example Fig.~1 of Ref.~\cite{Graf:2015tda}) but with a very large band going all the way to $P=0$. It is important to clarify that this band is not an error bar, but the envelope considering the different scenarios of strange quark mass proposed in \cite{Kurkela:2009gj} and a variation of the renormalization parameter $\bar{\Lambda}$ (see \cite{Fraga:2004gz,Kurkela:2016was,Gorda:2018gpy}, for more details).

Note that pQCD figures are usually shown normalized by the same case, same massless quark and chemical potentials conditions, only turning off the interactions (in Ref.~\cite{Graf:2015tda} the effect of masses are highlighted, but the chemical potential conditions are kept the same). Their intention is to show the changes in the equation of state due to the interactions. Our approach is different, and in our view complementary: we are only looking at free quarks, but we are normalizing our figures to different quark masses and, most importantly,  chemical potential values or conditions. Our intention is to measure how much these masses and chemical potentials affect the equation of state.

To do so, in this work, we have investigated the equation of state of a free gas of quarks focusing on how the conformal limit is reached when different chemical potentials are varied and different constraints (e.g. , for laboratory vs. astrophysics) are considered. This is done by using combinations of 1, 2, or 3 chemical potentials out of the 4 we consider, each related to a possible conserved quantity: baryon number B ($\mu_B$), electric charge ($\mu_Q$), strangeness ($\mu_S$), and lepton number ($\mu_\nu$). 
We have also derived expressions for massless quarks under different conditions and used the proportionality between the results to illustrate our discussion.

We have studied the effects of using different quark masses (including PDG values), number of flavors, and different ways to fix the various chemical potentials considered. The latter procedure implies enforcing charge neutrality and, when leptons were included, beta equilibrium. When leptons (electrons, muons, and their respective neutrinos) are present, the pressure in not altered. An exception is the case in which the lepton fraction is fixed. For different cases, we have quantified the deviation from the one-chemical potential (massless) conformal limit by verifying at which $\mu_B$ the pressure deviates by more than $10\%$. This value varied from $\mu_B=77$ to $48~897$ MeV. Depending of the values of chemical potentials, e.g, $\mu_Q$, even the light quark masses can become relevant at large $\mu_B$. This shows that one must be careful about making statements concerning comparisons with "the" conformal limit. 

Finally, we have shown that the behavior of the symmetry energy is monotonically increasing and does not depend on the light quark masses. 

\vspace{6pt} 

\section*{Author contributions}
Conceptualization, V.D.; methodology, V.D.; software, C.B. and V.D.; validation, R.B.J. and R.L.S.F.; formal analysis, V.D. and R.B.J.; investigation, C.B.. V.D., R.B.J. and R.L.S.F.; data curation, V.D. and R.B.J.; writing---original draft preparation, C.B. and V.D.; writing---review and editing, R.B.J. and R.L.S.F.; visualization, C.B. and V.D.; supervision, V.D. and R.L.S.F.; project administration, V.D.; funding acquisition, V.D. and R.L.S.F. All authors have read and agreed to the published version of the manuscript. 

\section*{Funding}
This research was funded by the National Science Foundation, grants number PHY1748621, MUSES OAC2103680, and~NP3M PHY-2116686 (V.D.). This work was partially supported by Conselho Nacional de Desenvolvimento Cient\'ifico 
e Tecno\-l\'o\-gico  (CNPq), grant number 312032/2023-4 (R.L.S.F.) and is also part of the project Instituto Nacional de Ci\^encia e Tecnologia - F\'isica Nuclear e Aplica\c{c}\~oes (INCT - FNA), grant number 464898/2014-5 (R.L.S.F.). It also received support from the Kent State University SURE program (C.B.).

\section*{Data availability}
The raw data supporting the conclusions of this article will be made available by the authors on request. 


\section*{Conflicts of interest}
The authors declare no conflicts of interest.

\section*{Abbreviations} 
The following abbreviations are used in this manuscript: perturbative Quantum Chromodynamics (pQCD), Particle Data Group (PDG).


\appendix
\section[\appendixname~\thesection]{Appendix: General Expressions}

For each particle $i$, we can write its chemical potential as the combination of the independent chemical potentials of the system (each associated with a conserved quantity) weighted by the respective particle quantum number. In our case, of conserved baryon number, electric charge, and strangeness, we have
\begin{equation}
\mu_i = Q_B \mu_B + Q_i \mu_Q + Q_{S_i} \mu_S,
\label{mu}
\end{equation}
where the baryon number for quarks is $Q_B=1/3$ and $Q_i$ and $Q_{S_i}$ are the electric charge and strangeness of each quark. Here $\mu_B$, $\mu_Q$, and $\mu_S$ are the baryon, electric charge, and strange chemical potentials of the system. In our formalism, the isospin chemical potential $\mu_I=\mu_Q$ \cite{Aryal:2020ocm}. Alternatively, Eq.~\ref{mu} can be derived from the principles of thermodynamics. One may start considering the additive property of the internal energy $U$ for a $n$-component system:
\begin{equation}
    U (\lambda S,\lambda V,\lambda N_1,... \lambda N_n) = \lambda  U (S,V,N_1,...N_n) \, , 
\end{equation}
where $\lambda$ is arbitrary and $S$, $V$ and $N_i$ denote the entropy, the volume and the number of particles of a given component $i$, respectively. Let us differentiate this “extensivity condition” with respect to $\lambda$:   
\begin{eqnarray}
    \frac{\partial U (\lambda S,...)}{\partial (\lambda S)} \frac{\partial (\lambda S)}{\partial \lambda} &+& \frac{\partial U (\lambda V,...)}{\partial (\lambda V)} \frac{\partial (\lambda V)}{\partial \lambda} \, + \, \frac{\partial U (\lambda N_1,...)}{\partial (\lambda N_1)} \frac{\partial (\lambda N_1)}{\partial \lambda} + ... \nonumber\\ &+& \frac{\partial U (\lambda N_n,...)}{\partial (\lambda N_n)} \frac{\partial (\lambda N_n)}{\partial \lambda} \, = \, U (S,V,N_1,...N_n) \, . 
\end{eqnarray}
Setting $\lambda=1$ in the above equation, we obtain:
\begin{equation}
    \frac{\partial U}{\partial S} S + \frac{\partial U}{\partial V} V + \frac{\partial U}{\partial N_1} N_1 + ... + \frac{\partial U}{\partial N_n} N_n = U \, . 
\end{equation}
Now, using the definition of the intensive parameters $T$ (temperature), $P$ (pressure) and $\mu_i$ (chemical potential), we arrive at the \emph{Euler equation}:
\begin{equation}
    U = TS - PV + \mu_1 N_1 + … + \mu_n N_n \, . 
\end{equation}
Inserting the above equations into the expression for the \emph{Gibbs energy} $G = U + PV - TS$, we have:
\begin{equation}
\label{GBrelation}
    G = TS - PV + \mu_1 N_1 + … + \mu_n N_n + PV - TS  \implies G = \sum_{i=1}^{n} \mu_i N_i \, , 
\end{equation}
which is known as the \emph{Gibbs-Duhem relation}. In the condition of chemical equilibrium, the Gibbs energy must be minimized with respect to one of the quantities $N_j$. For constant temperature and pressure, this condition reads:
\begin{equation}
\label{equilibrium}
\sum_{i=1}^{n} {\left( \frac{\partial G}{\partial N_i}
\right)}_{T,P,N_{i \neq j}} \frac{dN_i}{dN_j} = 0 \, .
\end{equation}
Now, let $\eta_j$ stand for the coefficient that gives the proportion of the component $j$ with respect to the other components of the system. If the component $j$ suffers a variation $dN_j = \bar{\eta_j}$, all the other components must also have a variation given by $dN_i = (\bar{\eta_j}/\eta_j) \eta_i$ in order to keep the balance between the components implied by the condition of chemical equilibrium. Therefore, $dN_i/dN_j = \eta_i/\eta_j$. Additionally, according to Eq.~\ref{GBrelation}, 
\begin{equation}
    \mu_i = {\left(  \frac{\partial G}{\partial N_i}
\right)}_{T,P,N_{i \neq j}} \, . 
\end{equation}
As a result, Eq.~\ref{equilibrium} can be written as:
\begin{equation}
\label{equilibrium2}
    \sum_{i=1}^{n} \eta_i \mu_i = 0 \, .
\end{equation}
Considering that baryon number $Q_B$, electric charge $Q$ and strangeness $Q_S$ are conserved quantities, the three conservation laws can be respectively written as
\begin{equation}
\sum_{i=1}^{n} \eta_i Q_{{B}_i} = 0 \, \, \textrm{,} \, \, \,
\, \, \sum_{i=1}^{n} \eta_i Q_{i} = 0 \, \, \, \, \, \textrm{and} \, \, \, \sum_{i=1}^{n} \eta_i Q_{{S}_i} = 0   \, .
\end{equation}
As long as we have $n$ variables and $3$ equations, it is possible to write $3$ of the $\eta_i$ as a functions of the other $n-3$, as follows:
\begin{equation}
\eta_1 Q_{{B}_1} + \eta_2 Q_{{B}_2} + \eta_3 Q_{{B}_3} = - \sum_{i \neq 1,2,3}^{n} \eta_i Q_{{B}_i}
\, ,
\end{equation}
\begin{equation}
\eta_1 Q_{1} + \eta_2 Q_{2} + \eta_3 Q_{3} = - \sum_{i \neq 1,2,3}^{n} \eta_i Q_{i}
\, ,
\end{equation}
\begin{equation}
\eta_1 Q_{{S}_1} + \eta_2 Q_{{S}_2} + \eta_3 Q_{{S}_3} = - \sum_{i \neq 1,2,3}^{n} \eta_i Q_{{S}_i}
\, .
\end{equation}
Clearly, the $3$ independent components chosen to construct the above relations are completely arbitrary. As a consequence, we may consider a certain species $1$ such that $Q_{{B}_1}=1$, $Q_{1}=0$ and $Q_{{S}_1}=0$; a certain species $2$ such that $Q_{{B}_2}=0$, $Q_{2}=1$ and $Q_{{S}_2}=0$; and a certain species $3$ such that $Q_{{B}_3}=0$, $Q_{3}=0$ and $Q_{{S}_3}=1$. In this case, the above equations simplify to
\begin{equation}
\label{independent}
\eta_1 = - \sum_{i \neq 1,2,3}^{n} \eta_i Q_{{B}_i}
\,   \textrm{,} \, \, \,
\, \, \eta_2 = - \sum_{i \neq 1,2,3}^{n} \eta_i Q_{i}
\, \, \, \,  \textrm{and} \, \, \,
\,  \eta_3 = - \sum_{i \neq 1,2,3}^{n} \eta_i Q_{{S}_i}
\, .
\end{equation}
Plugging Eq.~\ref{independent} into Eq.~\ref{equilibrium2}, we find:
\begin{equation}
    \sum_{i \neq 1,2,3}^{n} \eta_i \mu_i = \sum_{i \neq 1,2,3}^{n} \eta_i (Q_{{B}_i} \mu_1) + \sum_{i \neq 1,2,3}^{n} \eta_i (Q_{i} \mu_2) + \sum_{i \neq 1,2,3}^{n} \eta_i (Q_{{S}_i} \mu_3) \, . 
\end{equation}
Defining $\mu_1=\mu_B$ (the baryon chemical potential), $\mu_2=\mu_Q$ (the electric charge chemical potential) and $\mu_3=\mu_S$ (the strange chemical potential), the above equation can be rewritten as:
\begin{equation}
    \sum_{i \neq 1,2,3}^{n} \eta_i \mu_i = \sum_{i \neq 1,2,3}^{n} \eta_i (Q_{{B}_i} \mu_B) + \sum_{i \neq 1,2,3}^{n} \eta_i (Q_{i} \mu_Q) + \sum_{i \neq 1,2,3}^{n} \eta_i (Q_{{S}_i} \mu_S) \, . 
\end{equation}
Finally, since all the factors $\eta_i$ are independent, the above equation only holds if the coefficients are equal, i.e.,
\begin{equation}
\mu_i = Q_{{B}_i} \mu_B + Q_i \mu_Q + Q_{S_i} \mu_S \, ,
\end{equation}
which precisely corresponds to Eq.~\ref{mu}, if we consider that all quarks have an identical baryon number $Q_B$, such that $Q_{{B}_i}=Q_B=1/3$. 

The general expressions for energy density and pressure of a relativistic free Fermi gas of particles $i$ can be derived from the Dirac Lagrangian density extracting the diagonal components of the energy-momentum tensor (assuming an ideal fluid). The (number) density is simply the integral of the distribution function. Using the natural system of units, they are
\begin{equation}
n_i=\frac{g_i}{2\pi^2}\int_0^\infty dk_i~k_i^2(f_{_i+}-f_{i_-}),
\end{equation}
\begin{equation}
\varepsilon_i=\frac{g_i}{2\pi^2}\int_0^\infty dk_i E_i k_i^2(f_{i_+}+f_{i_-}),
\end{equation}
\begin{equation}
P_i=\frac{1}{3}\frac{g_i}{2\pi^2}\int_0^\infty dk_i\frac{k_i^4}{E_i}(f_{i_+}+f_{i_-}),
\end{equation}
where $g_i=6$ is the spin and color degeneracy factor, $k_i$ is the momentum,
\begin{equation}
E_i=\sqrt{k_i^2+m_i^2}\geq0 ,
\label{e}
\end{equation}
is the energy of the state, $m_i$ the mass,
$f_\pm$ the distribution function of particles and antiparticles
$f_{i_\pm}=(e^{(E_i\mp\mu_i)/T}+1)^{-1}$,
with $\mu_i$ being the particle chemical potential, and $T$ the temperature. 

In the $T=0$ limit, antiparticles provide no contribution, $f_-=0$, and $f_+=1$ up to the Fermi momentum, $k_i=k_{F_i}$, $E_i=\mu_i$ and the integrals for the above thermodynamic quantities are evaluated analytically
\begin{equation}
n_i=\frac{g_i}{6\pi^2}k_{F_i}^3,
\end{equation}
\begin{equation}
\varepsilon_i=\frac{g_i}{2\pi^2}\Bigg[\left(\frac{1}{8}m_i^2k_{F_i}+\frac{1}{4}k_{F_i}^3\right)\sqrt{m_i^2+k_{F_i}^2}-\frac{1}{8}m_i^4\ln{\frac{k_{F_i}+\sqrt{m_i^2+k_{F_i}^2}}{m_i}}\Bigg],
\end{equation}
\begin{equation}
P_i=\frac{1}{3}\frac{g_i}{2\pi^2}\Bigg[\left(\frac{1}{4}k_{F_i}^3-\frac{3}{8}m_i^2k_{F_i}\right)\sqrt{m_i^2+k_{F_i}^2}+\frac{3}{8}m_i^4\ln{\frac{k_{F_i}+\sqrt{m_i^2+k_{F_i}^2}}{m_i}}\Bigg].
\end{equation}

\section[\appendixname~\thesection]{Massless Quarks}
For the massless particle case, the expressions above further reduce to 
\begin{equation}
n_i=\frac{g_i}{6\pi^2}k_{F_i}^3=\frac{g_i}{6\pi^2}\mu_i^3,
\end{equation}
\begin{equation}
\varepsilon_i=\frac{g_i}{8\pi^2}k_{F_i}^4=\frac{g_i}{8\pi^2}\mu_i^4,
\end{equation}
\begin{equation}
P_i=\frac{1}{3}\frac{g_i}{8\pi^2}k_{F_i}^4=\frac{1}{3}\frac{g_i}{8\pi^2}\mu_i^4,
\label{pressure}
\end{equation}
reproducing $\varepsilon_i = 3P_i$.

Note that, in the case of massless free quarks, we can also write $\mu_i=k_i$. Therefore, we can write the chemical potential for each quark flavor using Eq.~\eqref{mu}
\begin{equation}
\mu_u = \frac 1 3 \mu_B + \frac 2 3 \mu_Q = k_u\ ,
\label{muu}
\end{equation}
\begin{equation}
\mu_d = \frac 1 3 \mu_B - \frac 1 3 \mu_Q = k_d\ , 
\label{mud}
\end{equation}
\begin{equation}
\mu_s = \frac 1 3 \mu_B - \frac{1}{3} \mu_Q + \mu_S = k_s\ .
\label{mus}
\end{equation}
\vspace{1mm}

We use the convention that both the strangeness and $\mu_S$ are positive. Alternatively, one could use both as negative without changing the results.
Eqs.~\ref{muu}, and \ref{mud} are equal if $\mu_Q=0$. Eqs.~\ref{muu}, \ref{mud}, and \ref{mus} are equal if $\mu_Q=0$ and $\mu_S=0$. The density and pressure of each quark flavor can be written further as
\begin{equation}
n_i = \frac{\mu_i^3}{\pi^2} = \frac{k_i^3}{\pi^2}\ , 
\end{equation}
\begin{equation}
P_i = \frac{\mu_i^4}{4\pi^2} = \frac{k_i^4}{4\pi^2} \ .
\end{equation}
Next, we discuss the pressure for specific conditions concerning number of flavors and chemical potential constraints (not including leptons):

\begin{itemize}
\item \underline{2-flavor, $\mu_Q = 0$}
\end{itemize}
\begin{equation}
P=P_u + P_d = 2P_u = \frac{2\mu_u^4}{4\pi^2} = \frac{\mu_B^4}{162\pi^2}= \frac{\mu_B^4}{1598.88}\ .
\label{17}
\end{equation}
%
\begin{itemize}
\item \underline{3-flavor, $\mu_Q = 0$, $\mu_S = 0$}
\end{itemize}
\begin{equation}
P=P_u+P_d+P_s=3P_u=\frac{3\mu_u^4}{4\pi^2}=\frac{\mu_B^4}{108\pi^2}=\frac{\mu_B^4}{1065.92}\ .
\end{equation}

\begin{itemize}
\item \underline{2-flavor, $\mu_Q$ fixed}
\end{itemize}
\begin{eqnarray}
P =P_u +P_d &=&\frac{1}{4\pi^2}\left(\mu_u^4+\mu_d^4 \right)=\frac{1}{4\pi^2}\left[\left(\frac{1}{3}\mu_B +\frac{2}{3}\mu_Q \right)^4 +\left(\frac{1}{3}\mu_B- \frac{1}{3}\mu_Q \right)^4 \right]\nonumber\\
&=& \frac{1}{4\pi^2} \Bigg(\frac{\mu_B^4}{81}+\frac{4\mu_B^3}{27}\frac{2\mu_Q}{3}+\frac{6\mu_B^2}{9} \frac{4\mu_Q^2}{9} +\frac{4\mu_B}{3} \frac{8\mu_Q^3}{27} +\frac{16\mu_Q^4}{81} \nonumber\\&+&\frac{\mu_B^4}{81} -\frac{4\mu_B^3}{27} \frac{\mu_Q}{3} + \frac{6\mu_B^2}{9} \frac{\mu_Q^2}{9} -\frac{4 \mu_B}{3} \frac{\mu_Q^3}{27} +\frac{\mu_Q^4}{81} \Bigg)\nonumber\\
&=& \frac{1}{324\pi ^2}\left[2\mu_B^4+4\mu_B^3 \mu_Q +30\mu_B^2 \mu_Q^2 +28\mu_B \mu_Q^3 +17\mu_Q^4 \right]\ .
\label{21}
\end{eqnarray}
\vspace{1mm}

\begin{itemize}
\item \underline{2-flavor, $\mu_Q$ from charge neutrality}
\end{itemize}

Starting from $\sum_i Q_i n_i = 0$
\begin{equation}
\frac{2}{3} n_u - \frac{1}{3} n_d = 0\ , 
\end{equation}
\begin{equation}
\frac 2 3 \frac{\mu_u^3}{\pi^2} - \frac 1 3 \frac{\mu_d^3}{\pi^2} = 0\ ,
\end{equation}
\begin{equation}
2\mu_u^3=\mu_d^3\ ,
\label{eqmus}
\end{equation}
\begin{equation}
2\left(\frac{1}{3} \mu_B + \frac{2}{3} \mu_Q \right)^3 = \left(\frac{1}{3} \mu_B - \frac{1}{3} \mu_Q  \right)^3\ ,
\end{equation}
\begin{equation}
2^\frac{1}{3}\frac{1}{3} \mu_B - \frac{1}{3} \mu_B = - 2^\frac{1}{3}\frac{2}{3} \mu_Q - \frac{1}{3}\mu_Q\ ,
\end{equation}
\begin{equation}
\mu_Q = \frac{-\left(2^\frac{1}{3}-1 \right) \mu_B}{2^\frac{4}{3}+1}= -0.07\ \mu_B\ .
\label{eqmus2}
\end{equation}
We can then use Eqs.~\ref{eqmus} and \ref{eqmus2} to calculate the pressure
\begin{eqnarray}
P&=&P_u + P_d = \frac{1}{4 \pi^2} \left(\mu_u ^4 + \mu_d ^4 \right) = \frac{1}{4\pi^2}\left(\mu_u^4 + 2^\frac{4}{3} \mu_u^4 \right)\nonumber\\
&=&\frac{1}{4\pi^2} \left(1 + 2^\frac{4}{3}\right) \mu_u^4 = \frac{1}{4\pi^2} \left(1+2^\frac{4}{3} \right) \left(\frac{1}{3} \mu_B + \frac{2}{3} \mu_Q \right)^4\nonumber\\
&=&\frac{1}{4\pi^2}\left(1+2^\frac{4}{3}\right) \left[\frac{1}{3} \mu_B - \frac{2}{3} \left(\frac{2^\frac{1}{3}-1}{2^\frac{4}{3}+1} \mu_B \right)\right]^4\nonumber\\
&=&\frac{1}{4\pi^2} \left(1+2^\frac{4}{3} \right) \left[\frac{2^\frac{4}{3}+1-2^\frac{4}{3}+2}{3(2^\frac{4}{3}+1)}\right]^4 \mu_B^4\nonumber\\
&=&\frac{1}{4\pi^2 (2^\frac{4}{3}+1)^3}\mu_B^4 = \frac{\mu_B^4}{1721.59}\ .
\label{32}
\end{eqnarray}

\begin{itemize}
    \item \underline{3-flavor, $\mu_Q$ fixed, $\mu_S = 0$}
\end{itemize}
\begin{equation}
P = P_u + P_d + P_s = \frac{1}{4\pi^2} (\mu_u^4 + \mu_d^4 + \mu_s^4) = \frac{1}{4\pi^2}(\mu_u^4 + 2\mu_d^4)\ ,
\end{equation}
because $\mu_d=\mu_s$ are equal, resulting in
\begin{eqnarray}
P &=&\frac{1}{4\pi^2}\left[\left(\frac{1}{3}\mu_B +\frac{2}{3}\mu_Q \right)^4 +2\left(\frac{1}{3}\mu_B- \frac{1}{3}\mu_Q \right)^4 \right]\nonumber\\
&=& \frac{1}{4\pi^2} \Bigg(\frac{\mu_B^4}{81}+\frac{4\mu_B^3}{27}\frac{2\mu_Q}{3}+\frac{6\mu_B^2}{9} \frac{4\mu_Q^2}{9} +\frac{4\mu_B}{3} \frac{8\mu_Q^3}{27} +\frac{16\mu_Q^4}{81} \nonumber\\
&+&2\frac{\mu_B^4}{81} -2\frac{4\mu_B^3}{27} \frac{\mu_Q}{3} + 2\frac{6\mu_B^2}{9} \frac{\mu_Q^2}{9} -2\frac{4 \mu_B}{3} \frac{\mu_Q^3}{27} +2\frac{\mu_Q^4}{81} \Bigg)\nonumber\\
&=&\frac{1}{324\pi^2} \left(3\mu_B^4 + 36\mu_B^2 \mu_Q^2 + 24\mu_B \mu_Q^3 + 18\mu_Q^4 \right)\ .
\label{previous}
\end{eqnarray}

\begin{itemize}
\item \underline{3-flavor, $\mu_Q$ from charge neutrality, $\mu_S=0$}
\end{itemize}

Starting again from $\sum_i Q_i n_i = 0$
\begin{equation}
\label{a37}
\frac{2}{3}n_u-\frac{1}{3}n_d-\frac{1}{3}n_s=0\ ,
\end{equation}
\begin{equation}
\frac 2 3 \left(\frac{\mu_u^3}{\pi^2} \right)-\frac 1 3 \left(\frac{\mu_d^3}{\pi^2} \right) \frac 1 3 -\left(\frac{\mu_s^3}{\pi^2} \right) = 0\ ,
\end{equation}
\begin{equation}
2\mu_u^3-\mu_d^3-\mu_s^3=0\ ,
\end{equation}
but, since in this case $\mu_d=\mu_s$, we have:
\begin{equation}
\label{a40}
\mu_u^3= \mu_d^3\ ,
\end{equation}
which implies (from Eqs.~\ref{muu} and \ref{mud}) $\mu_Q =0$ and reproduces the 3-flavor case with $\mu_Q=0$, $\mu_S=0$.

\begin{itemize}
\item \underline{3-flavor, zero net strangeness}
\end{itemize}

Starting from $\sum Q_{S_i}n_i=0$, at $T=0$ this implies $n_s=0$, no matter if $\mu_Q =0$ or $\mu_Q \neq 0$. As a consequence, this case reproduces the respective 2-flavor case.

\begin{itemize}
\item \underline{3-flavor, $\mu_Q$ fixed, $\mu_S$ fixed}
\end{itemize}
\begin{eqnarray}
P &=& P_u + P_d + P_s = \frac{1}{4\pi^2} (\mu_u^4 + \mu_d^4 + \mu_s^4)\nonumber\\
 &=& \frac{1}{4\pi^2}\left[\left(\frac{1}{3}\mu_B +\frac{2}{3}\mu_Q \right)^4 +\left(\frac{1}{3}\mu_B- \frac{1}{3}\mu_Q \right)^4 + \left(\frac{1}{3}\mu_B- \frac{1}{3}\mu_Q +\mu_S\right)^4\right]\ .
\end{eqnarray}
Using the result from Eq.~\ref{previous}
\begin{eqnarray}
\label{grande}
P&=&\frac{1}{324\pi^2} \left(3\mu_B^4+36\mu_B^2\mu_Q^2+24\mu_B\mu_Q^3+18\mu_Q^4 \right)\nonumber\\
&+& \frac{1}{4\pi^2}\Bigg(\mu_S^4 - \frac {4} {27} \mu_Q^3 \mu_S - \frac 4 3 \mu_Q \mu_S^3 + \frac 4 3 \mu_B \mu_S^3 +\frac {4} {27} \mu_B^3 \mu_S \nonumber\\ 
&+& \frac 6 9 \mu_Q^2 \mu_S^2 + \frac 6 9 \mu_B^2 \mu_S^2 - \frac {12} {27} \mu_B^2 \mu_Q \mu_S + \frac {12} {27} \mu_B \mu_Q^2 \mu_S - \frac {12} {9} \mu_B \mu_Q \mu_S^2 \Bigg)\ .
\end{eqnarray}

\begin{itemize}
\item \underline{3-flavor $\mu_Q=0 , \mu_S $ fixed}
\end{itemize}
Using Eq.~\ref{grande} with $\mu_Q=0$
\begin{equation}
P =\frac{1}{\pi^2}\left(\frac{1}{108} \mu_B^4 +\frac{\mu_S ^4}{4}+\frac{1}{3}\mu_B \mu_s ^3 +\frac{1}{27}\mu_B ^3 \mu_S +\frac{1}{6} \mu_B ^2 \mu_S ^2 \right)\ .
\end{equation}

\begin{itemize}
\item \underline{3-flavor, $\mu_Q$ from charge neutrality, $\mu_S$ fixed}
\end{itemize}
Starting from $\sum Q_i n_i=0$
\begin{equation}
\frac{2}{3}n_u -\frac{1}{3}n_d -\frac{1}{3}n_s = 0\ ,
\end{equation}
\begin{equation}
2 \mu_u ^3- \mu_d ^3 -\mu_s ^3=0\ ,
\end{equation}
\begin{equation}
2 \left( \frac{1}{3}\mu_B +\frac{2}{3}\mu_Q \right)^3 - \left( \frac{1}{3}\mu_B -\frac{1}{3}\mu_Q \right)^3 - \left( \frac{1}{3}\mu_B -\frac{1}{3}\mu_Q +\mu_S \right)^3=0\ ,
\end{equation}
\begin{eqnarray}
&&\frac{2\mu_B^3}{27}+\frac{12 \mu_B ^2 \mu_Q}{27} +  \frac{24 \mu_B \mu_Q^2}{27} +\frac{16\mu_Q^3}{27} -\frac{2\mu_B^3}{27}+ \frac{6\mu_B ^2 \mu_Q}{27} - \frac{6\mu_B \mu_Q ^2}{27}  \nonumber
\\ 
&+& \frac{2 \mu_Q ^3}{27}-\mu_S ^3 -\frac{3 \mu_B ^2 \mu_S}{9} +\frac{6\mu_B \mu_Q \mu_S}{9} - \frac{3\mu_Q^2 \mu_S}{9} -\frac{3\mu_B \mu_S^2}{3}+\frac{3}{3}\mu_Q \mu_S^2=0\ ,
\end{eqnarray}
\begin{equation}
\frac{2 \mu_B ^2 \mu_Q}{3} +  \frac{2 \mu_B \mu_Q^2}{3} +\frac{2\mu_Q^3}{3} -\mu_S^3-\frac{\mu_B^2\mu_S}{3} +\frac{2\mu_B \mu_Q \mu_S}{3} - \frac{\mu_Q^2 \mu_S}{3} -\mu_B \mu_S^2+\mu_Q \mu_S^2=0\ .
\end{equation}
In the above expression, we still need to isolate $\mu_Q$ and replace in Eq.~\ref{grande}.

\bibliographystyle{unsrtnat}
\bibliography{ConformalLimit}

\end{document}